\documentclass[twocolumn,amsmath,amssymb]{revtex4}
\usepackage[ansinew]{inputenc}
\usepackage{graphicx}
\usepackage{pst-all}
\usepackage{color}
\usepackage{dcolumn}
\usepackage{amsmath}
\usepackage{amsthm}
\usepackage{bm}
\usepackage{layout}
\usepackage{float}
\usepackage{txfonts}
\usepackage{amsfonts}
\usepackage{amssymb}%
\usepackage[ansinew]{inputenc}
\usepackage{graphicx}
\usepackage{amsmath}
\usepackage{amsthm}
\usepackage{bm}
\usepackage{layout}
\usepackage{float}
\usepackage{amsfonts}
\usepackage{amssymb}
\usepackage{array}
\usepackage{multirow,bigdelim}
\usepackage{enumitem}
\usepackage{epstopdf}
\usepackage{CJK}
\usepackage{longtable}
\usepackage{amsmath,amssymb,bm}
\usepackage{graphicx,color}
\usepackage[squaren]{SIunits}
\usepackage[english]{babel}
\usepackage{amstext}
\usepackage{amsthm}
\usepackage{latexsym}
\usepackage{array}
\usepackage{color}
\usepackage{float}
\usepackage{microtype}

\usepackage{multirow}
\usepackage{dcolumn}
\usepackage{soul}
\usepackage{bm}
\usepackage{graphicx}

\usepackage[colorlinks=true,linkcolor=blue,
pagecolor=blue,filecolor=blue, menucolor=blue,urlcolor=blue,
citecolor=blue,anchorcolor=blue]{hyperref}%

\begin{document}
\title{Underlying topological Dirac nodal line mechanism of
anomalously large electron-phonon coupling strength on Be (0001)
surface}

\author{Ronghan Li$^{1,2}$}

\author{Jiangxu Li$^{1,2}$}
\author{Lei Wang$^{1,2}$}
\author{Hui Ma$^{1}$}

\author{Dianzhong Li$^{1}$}

\author{Yiyi Li$^{1}$}

\author{Xing-Qiu Chen$^{1}$}
\email[Corresponding author:
]{xingqiu.chen@imr.ac.cn}
\affiliation{$^{1}$Shenyang National Laboratory
for Materials Science, Institute of Metal Research, Chinese Academy
of Science, 110016 Shenyang, Liaoning, China}
\affiliation{$^{2}$ School of Materials Science and
 Engineering, University of Science and Technology of China, Heifei, P. R. China}
\date{\today}

\begin{abstract}

Beryllium was recently discovered to harbor a Dirac nodal line (DNL)
in its bulk phase and the DNL-induced non-trivial drumhead-like
surface states (DNSSs) on its (0001) surface, rationalizing several
already-existing historic puzzles [Phys. Rev. Lett., \textbf{117},
096401 (2016)]. However, to date the underlying mechanism, as to why
its (0001) surface exhibits an anomalously large electron-phonon
coupling effect ($\lambda_{e-ph}^s$ $\approx$ 1.0), remains
unresolved. Here, by means of first-principles calculations we have
evidenced that the coupling of the DNSSs with the phononic states
mainly contributes to its novel surface \emph{e-ph} enhancement.
Besides that the experimentally observed $\lambda_{e-ph}^s$ and the
main Eliashberg coupling function (ECF) peaks have been reproduced
well, we have decomposed the ECF,
$\alpha^{2}$$F$(\emph{k},\textbf{\emph{q}};\emph{v}), and the
\emph{e-ph} coupling strength
$\lambda(\emph{k},\textbf{\emph{q}};\emph{v})$ as a function of each
electron momentum (\emph{k}), each phonon momentum
(\textbf{\emph{q}}) and each phonon mode ($v$), evidencing the
robust connection between the DNSSs and both
$\alpha^{2}$$F$(\emph{k},\textbf{\emph{q}};\emph{v}) and
$\lambda(\emph{k},\textbf{\emph{q}};\emph{v})$. The results reveal
the strong \emph{e-ph} coupling between the DNSSs and the phonon
modes, which contributes over 80$\%$ of the $\lambda_{e-ph}^s$
coefficient on the Be (0001) surface. It highlights that the
anomalously large \emph{e-ph} coefficient on the Be (0001) surface
can be attributed to the presence of its DNL-induced DNSSs,
clarifying the long-term debated mechanism.
\end{abstract}


\maketitle

In difference from both topological Dirac semimetals and topological
Weyl semimetals which host isolated Dirac cones and Weyl nodes in
their bulk phases, the class of topological Dirac nodal line
semimetals (DNLs)
~\cite{TL-1,TL-2,TL-3,be0,dnl1,dnl2,dnl3,dnl4,dnl5,dnl6,dnl7,dnl8,
dnl9,dnl10,dnl11,dnl12,dnl13,dnl14,dnl15,dnl16,dnl17,
dnl18,dnl19,dnl20,dnl21,dnl22,dnl23, dnl24,dnl25,dnl26,dnl27,dnl30,
dnl28,dnl29,dnl31,dnl32,nr7,nr8,nr1,nr2,nr3,nr4,nr5,nc1,nc2,nc3,jxli2018,
nl1,nk1,nls1,nls2,40,41,42,43,44,46} exhibit the fully closed lines
around the Fermi level due to the continuously linear crossings of
the bulk energy bands in the lattice momentum space. The projection
of the DNLs onto a certain surface would result in a closed ring, in
which the topologically protected nearly-flat drumhead-like
non-trivial surface states (DNSSs) occur. This kind of exotic band
structures render various novel properties. Besides the common
DNSSs-induced high electronic density around the Fermi energy on the
surface, there are still giant surface Friedel
oscillation~\cite{be0}, flat Landau level~\cite{40}, long-range
Coulomb interaction ~\cite{41}, special collective
modes~\cite{dnl21}, flat optical conductivity
~\cite{dnl25,dnl26,dnl27,dnl28}, giant magnetoresistance and
mobility~\cite{dnl24}, and unconventional enhancement of effective
mass~\cite{dnl29} as well as a potential route to achieve
high-temperature superconductivity ~\cite{dnl18,dnl19,dnl20} and
catalytic candidates~\cite{jxli2018}, and so on.

Recently, the discoveries of the DNL in the pure beryllium
metal~\cite{be0} and this DNL-induced robust DNSSs on its (0001)
surface rationalize three already-existing historic puzzles
~\cite{be1,be2,exp,be3,be8,be6,be7,be9} of ($i$) the long-standing
question of the surface states observed by the angle-resolved
photoemission (ARPES) experiments, ($ii$) the underlying physics of
the severe deviations of its surface electronic structures from the
description of the nearly free electron picture, and ($iii$) the
substantial mechanism of the giant Friedel oscillations on the
(0001) surface. Although these three puzzles have been resolved, the
origin of its anomalously large electron-phonon (\emph{e-ph})
coupling effect on the (0001) surface still remains opening. As
early as in the 1998, the \emph{e-ph} coupling strength,
$\lambda_{e-ph}^s$, on the Be (0001) surface was measured to be 1.15
$\pm$ 0.1 or 1.18 $\pm$ 0.07 by the ARPES
experiment~\cite{besurf1,be4}, being about five times the bulk Be
value $\lambda_{e-ph}^b$ = 0.24~\cite{bebulk,Mahan2000}. On basis of
these experimental measurements, the authors claimed that the
surface superconductivity with a potential high critical temperature
may exist on the Be (0001) surface ~\cite{besurf1}. The subsequent
ARPES reports on the \emph{e-ph} coupling strength
$\lambda_{e-ph}^s$ was in a range from 0.6 to
1.18~\cite{be2000,be2004,besurf1,be4,be2003,exp_ph,be1997}, mainly
because of the anisotropic \emph{e-ph} interaction on the Be (0001)
$\overline{\Gamma}$ surface states~\cite{be2009}. However, the
reported low $\lambda_{e-ph}^s$ = 0.6~\cite{be2004} was
experimentally found to be indeed caused by surface oxygen
contamination~\cite{be2009}. Although \emph{ab initio} calculations
for Be surface reproduced both the measured geometric and electronic
structures as well as surface phonon dispersions, the known
calculations for the \emph{e-ph} coupling on Be (0001) surface did
not reproduce the measurements~\cite{be2009,be5}. The early
DFT-derived $\lambda_{e-ph}^s$ is 0.90~\cite{be2003,be2006}, but the
subsequent work in Ref. \cite{be2009} stated that the previously
calculated value was wrong due to the programming error. The latest
ARPES experiments \cite{be5} even claimed that none of the main
experimental Eliashberg coupling function (ECF) peaks in the
low-frequency range was captured by the previous
calculations~\cite{be2003,be2006}. Therefore, the mechanism
dramatically triggering the anomalously large \emph{e-ph} coupling
strength of the Be (0001) surface remains opening and the
discrepancies between experimental measurement and theoretical
calculations would also need to be urgently clarified.

Returning to the fact that bulk Be exhibits the novel DNL and the
DNL-induced DNSSs on its (0001) surface~\cite{be0}, we have strongly
suspected whether or not the highly localized and nearly flat DNSSs
result in the occurrence of the anomalously large \emph{e-ph}
coupling strength. Here, through first-principles calculations
within the framework of Density Function Theory
(DFT)\cite{dft1,dft2} by employing the Quantum Espresso~\cite{qe}
(QE) code with the norm-conserving pseudopotential
~\cite{pz}(details refer to supplementary method~\cite{SM}), we have
revisited the problem of the \emph{e-ph} coupling strength of the Be
(0001) surface by establishing the connection of both the ECF,
$\alpha^{2}$$F$(\emph{k},\textbf{\emph{q}};\emph{v}), and the
\emph{e-ph} coupling strength,
$\lambda(\emph{k},\textbf{\emph{q}};\emph{v})$. Remarkably, we have
evidenced the strong \emph{e-ph} coupling of the DNSSs with the
phonon modes, contributing over 80\% of the $\lambda^s_{e-ph}$ of
the Be (0001) surface. Accordingly, these results confirm that the
anomalously large \emph{e-ph} coupling strength on the Be (0001)
surface is substantially ascribed to the topologically protected
DNL-induced DNSSs.

\begin{figure}[hbt]
\centering
\includegraphics[width=0.49\textwidth]{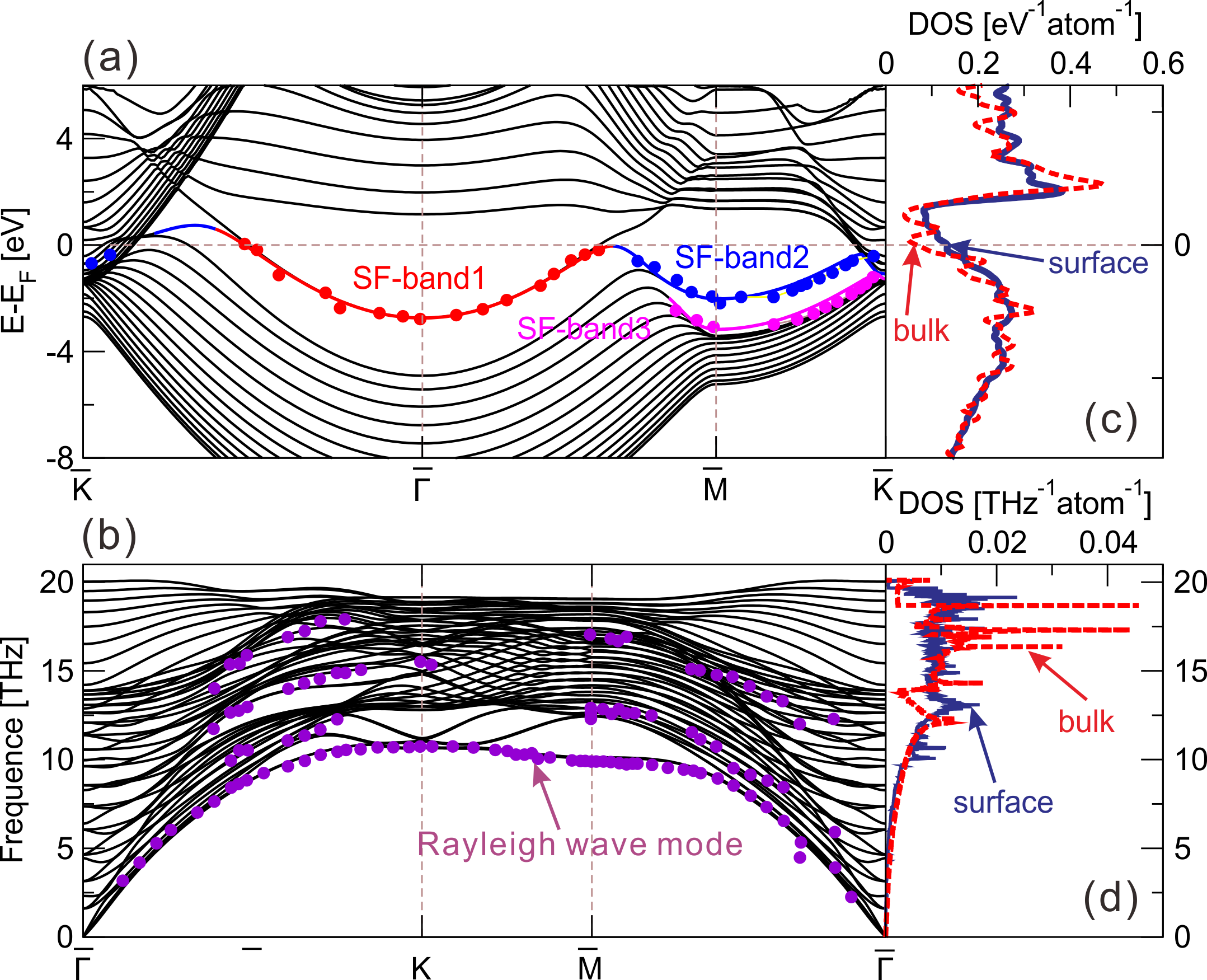}
\caption{DFT-derived electronic band structure (a,c) and phonon
dispersion (b,d) of the (0001) surface in the hcp Be metal. Panel
(a): the surface electronic band structures along the high-symmetric
lines as compared with available ARPES experimental data. Panel (b):
the surface phonon dispersion along the high-symmetric lines as
compared with available experimentally observed surface phonon
dispersions. Panel (c and d): the surface electronic (c) and phonon
(d) densities of states in comparison with the total density of
states of its bulk phase.} \label{fig1}
\end{figure}

\begin{figure}[hbt]
\centering
\includegraphics[width=0.49\textwidth]{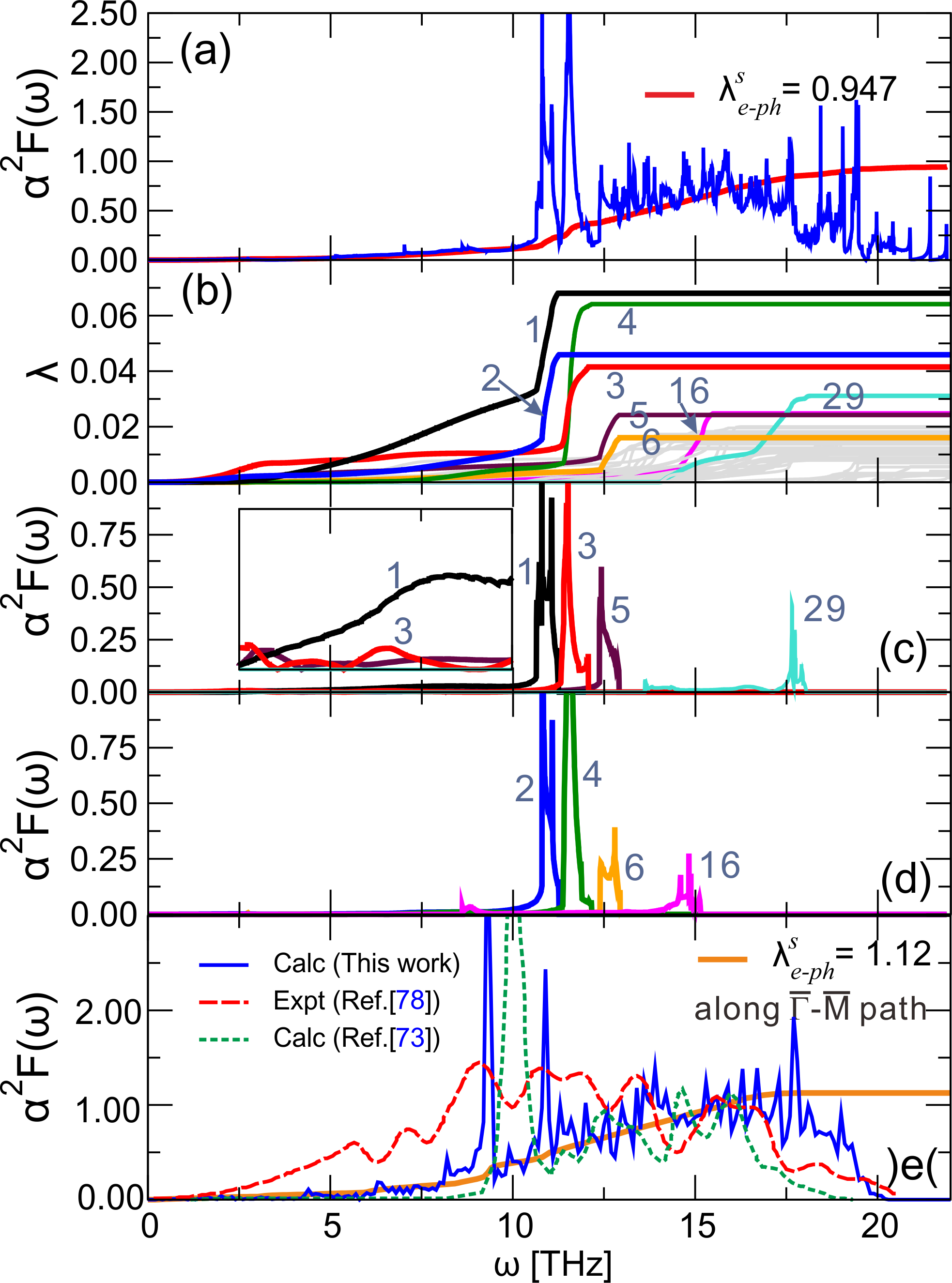}
\caption{DFT-derived Eliashberg coupling function (ECF),
$\alpha^2F(\omega)$, and the coupling strength, $\lambda$, of the Be
(0001) surface. Panel (a): the total ECF and the total \emph{e-ph}
coupling strength; Panel (b): 48 mode-resolved \emph{e-ph} coupling
strengths, $\lambda$. The $\lambda$ values of No.1 $\sim$ No.6,
No.16 and No.29 modes are highlighted and the other 40 modes are
plot in a grey background; Panel (c and d): the mode-resolved ECFs
for eight selected phonon modes [No.1 (RW mode), No.3, No.5 and
No.29 modes in panel (c); No.2, No.4, No.6 and No.16 modes in panel
(d)]; Panel (e): the derived ECF along the $\bar{\Gamma}$ -
$\bar{M}$ direction as compared with the ARPES-extracted data
multiplied with a factor of 3 for easy comparison \cite{be5} and
previous calculated data~\cite{be2006}.} \label{fig2}
\end{figure}

The current calculations of the electronic band structures have
reproduced well our previously published results ~\cite{be0}. The
DNL projection onto the (0001) surface exhibits a closed ring
surrounding $\bar{\Gamma}$ in which the topologically protected
DNSSs appear~\cite{be0}. Along the
$\bar{K}$-$\bar{\Gamma}$-$\bar{M}$ paths in the surface BZ, these
DNSSs [see SF-band1 in Fig.~\ref{fig1}(a)] disperse parabolically
around the centered $\bar{\Gamma}$ point, in nice agreement with the
experimental findings obtained by ARPES\cite{be1,exp}.
In addition, we have derived the surface phonon dispersions, as
shown in Fig.~\ref{fig1}(b and d). Our calculations are in a nice
agreement with the experimentally observed dispersions
\cite{be1,be2,be2000,be2004,be2009,be3} and the previous
calculations \cite{be2003,ph1_calc}. Particularly, it can be seen
that the highly localized surface phonon mode [Reyleigh waves (RW)
mode~\cite{be1997}] in Fig.~\ref{fig2}(b), dominated by the
vibration of the topmost atom along the surface normal, is soft and
very sensitive to the interplanar spacing between the first and
second topmost atomic layers. In comparison with its bulk phase, the
Be (0001) surface shows the apparent differences: (\emph{i}) the
surface electronic density at the Fermi level is much larger than
that of the bulk phase due to the presence of the DNSSs
(Fig.~\ref{fig1}c), and (\emph{ii}) the two extra peaks at 10 THz
and 11 THz occur in the surface phonon density and the apparent
large peak is shift to a higher frequency of 12.5 THz with respect
to that of its bulk phase (Fig.~\ref{fig1}d).

We have theoretically derived the ECF,
$\alpha^{2}$$F$$($$\omega$$)$, and the \emph{e-ph} coupling
strength, $\lambda$, for both the bulk phase and the (0001) surface
[see supplementary Fig. S1~\cite{SM} and Fig.~\ref{fig2}]. The
obtained  $\lambda_{e-ph}^b$ = 0.254 for the bulk phase
(supplementary Fig. S1~\cite{SM}) and $\lambda_{e-ph}^s$ =  0.947
for the Be (0001) surface [Fig.~\ref{fig2}(a)], in the nice
agreement with the experimental data (0.24 for bulk phase
\cite{Mahan2000,bebulk}; 1.15 $\pm$ 0.1 ~\cite{besurf1} and 1.18
$\pm$ 0.07 ~\cite{be4} for the Be (0001) surface). The previous
calculations \cite{be2003,be2006} did not correctly reproduced the
the frequency ranges of the experimental ECF peaks \cite{be2009,be5}
in the low-frequency region along the specified $\bar{\Gamma}$ -
$\bar{M}$ direction in the Be (0001) surface. As compared in
Fig.~\ref{fig2}(e), the conspicuous difference in ECF is the large
peak in the previous theoretical calculation~\cite{be2003,be2006} at
about 10 THz, on which the experimental data exhibited a valley
\cite{be5}. Importantly, along the $\bar{\Gamma}$ - $\bar{M}$
direction our current calculations reproduced well these
experimentally observed positions of ECF peaks in
Fig.~\ref{fig2}(e). Furthermore, in order to elucidate whether the
electronic DNSSs have effects on the \emph{e-ph} coupling strength
on the Be (0001) surface, we need to derive the ECF
$\alpha^{2}$$F$($\omega$) at the \emph{k} momenta where the DNSSs
would exist in the surface BZ. Therefore, we have decomposed the
total ECF $\alpha^{2}$$F$$($$\omega$$)$ \cite{theory} into the
function of each electron momentum ($k$), each phonon momentum
(\textbf{\emph{q}}) and each phonon mode ($v$) in the (0001) surface
BZ as,
\begin{equation}\label{equ4}
\alpha^{2}F(\emph{k},\emph{\textbf{q}};v) = \sum_{\emph{i,f}}
|g_{\textbf{\emph{q}},v}(\emph{k},\emph{i},\emph{f})|^2 \delta
(\epsilon_f - \epsilon_i \mp \omega_{\textbf{\emph{q}},v}),
\end{equation}
where the term of
$g_{\textbf{\emph{q}},v}(\emph{k},\emph{i},\emph{f})$ is the
so-called \emph{e-ph} matrix element, which represents the
probability of electron scattering from an initial electron state
\emph{i} with a momentum \textbf{\emph{k}} to a final electron state
\emph{f} interacted by a phonon with a momentum \textbf{\emph{q}}
and a mode index $v$. This term can be derived as follows,
\begin{equation}\label{equ2}
g_{\textbf{\emph{q}},v}(\emph{k},i,f) =
\sqrt{\frac{\hbar}{2M\omega_{\textbf{\emph{q}},v}}}\langle
\varphi_{i,\emph{k}}|\delta
V^{\textbf{\emph{q}},v}_{scf}|\varphi_{f,\emph{k} \pm
\textbf{\emph{q}}} \rangle,
\end{equation}
where $M$ is the atom mass and
$\delta$$V^{\textbf{\emph{q}},v}_{scf}$ is the gradient of the
self-consistent potential of the atomic displacements induced by the
phonon mode $v$ with a momentum \textbf{\emph{q}}\cite{theory}. In
addition, the total \emph{e-ph} coupling strength can be decomposed
into $\lambda(\emph{\textbf{q}};v)$ of each phonon momentum
(\textbf{\emph{q}}) and each phonon mode ($v$) over all electron
\emph{k} momenta in the (0001) surface BZ as,
\begin{equation}\label{equ5}
\lambda (\emph{\textbf{q}};v) = 2\int
\frac{dk}{\Omega_{BZ}}\frac{\alpha^{2}F(\emph{k},
\emph{\textbf{q}};v)}{N(e_F)\hbar\omega_{\textbf{q},v}},
\end{equation}
where $N(e_F)$ is the electronic density of states at the Fermi
level and $\Omega_{BZ}$ is the area of the surface BZ. In terms of
Equ.~(\ref{equ5}) we have first calculated mode-resolved $\lambda$
over all electron \emph{k} and phonon $q$ momenta in
Fig.~\ref{fig2}(b) to elucidate the effects of phonon modes on
$\lambda_{e-ph}^s$. Among them, the No.1 RW mode contributes the
largest $\lambda$ value of 0.068. In combining both the
mode-resolved ECF $\alpha^2F$($\omega$) in Fig.~\ref{fig2}c and
\emph{e-ph} coupling strength, $\lambda$ in Fig.~\ref{fig2}b, it can
be clearly referred that the $\lambda$ of the soft No.1 RW mode is
mainly originated from two parts. The first part is from the broad
ECF peak [see the enlarged inset in Fig.~\ref{fig2}(c)] in a range
from 5 THz $\sim$ 10 THz, in which the ECF along the
$\bar{\Gamma}$-$\bar{M}$ direction makes the great contribution
[Fig.~\ref{fig2}(e)]. The second part is from the sharp ECF peak in
a range of 10.3 THz $\sim$ 11.2 THz for the No. 1 RW mode. In very
similarity to the No.1 RW mode, the No.2 mode results in a sharp ECF
peak, almost overlapping with the second sharp ECF peak of the No.1
mode in the nearly same frequency region [Fig.~\ref{fig2}(a,b,c)].
It also contributes the third largest $\lambda$ value of 0.046. The
second sharp peak in the total ECF in the frequency range from 11.3
THz to 11.9 THz [Fig. ~\ref{fig2}(a)] is fully comprised by the
phonon No.3 and No. 4 modes, which mainly exhibit in-planed
vibration of the first topmost atomic layer in coupling with the
small out-planar vibration of the second topmost atomic layer along
the surface normal. They contribute the fourth and second largest
$\lambda$ of 0.044 and 0.064, respectively. With increasing the
frequency to the region from 12.4 THz to 12.9 THz the nearly
overlapped ECF peaks occur from both the phonon No.5 and No.6 modes
[Fig. ~\ref{fig2}(b,c,d)], which highlight the atomic vibrations of
the first and third topmost atomic layers along the surface normal.
They contribute to the relatively large $\lambda$ values of 0.024
and 0.016, respectively. Both No. 16 and No. 29 modes also
contribute a comparably high $\lambda$ of 0.024 and 0.032 in the
high-frequency region. Although all other phonon modes contribute a
sizable \emph{e-ph} coupling strength, their contributed $\lambda$
values are in general smaller than those of the above eight modes
[Fig. ~\ref{fig2}(b)]. Therefore, the six modes (No.1 to No.6) in
the low-frequency region and the two modes (No.16 and No.29) in the
high-frequency region dominantly contribute to the \emph{e-ph}
coupling strength. Interestingly, we have even recognized that these
eight modes exhibit highly localized surface phonon states
[Fig.~\ref{fig1}(b)].

\begin{figure*}[hbt]
\centering
\includegraphics[width=0.95\textwidth]{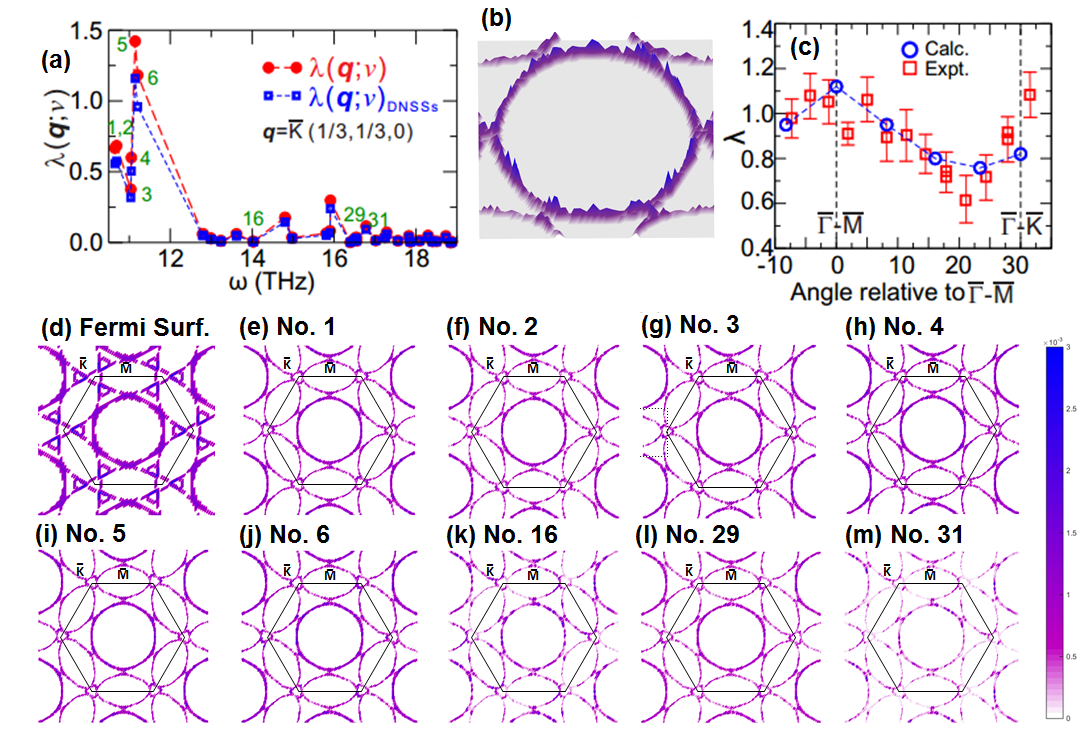}
\caption{The (\textbf{\emph{q}};$v$)-resolved $\lambda$ and ECFs at
the phonon $\bar{K}$ momentum \textbf{\emph{q}} =  (1/3, 1/3,
0)~\cite{qk} on the Be (0001) surface. Panel (a), The
(\textbf{\emph{q}};$v$)-resolved $\lambda$ and the number denotes
the corresponding vibration mode (the dashed lines are just a guide
to the eye). Panel (b), the 3D plot of the ECF of the No.1 mode
shows a strong anisotropic \emph{k}-dependent feature. Panel (c),
anisotropic $\lambda$ along the directions rotating from the
$\bar{\Gamma}$-$\bar{M}$ to $\bar{\Gamma}$-$\bar{K}$ by varying the
angle relative to the $\bar{\Gamma}$-$\bar{M}$ direction. Panel (d),
The derived electronic Fermi surface of the Be (0001) surface.
Panels (e) to (m), the (\textbf{\emph{q}};$v$)-resolved ECFs of No.1
$\sim$ No.6, No.16, No.29, and No.31 modes.} \label{fig3}
\end{figure*}

In order to understand in-depth the \emph{e-ph} coupling between the
DNSSs and both the ECF and $\lambda_{e-ph}^s$, we have derived the
(\textbf{\emph{q}};\emph{v})-resolved
$\lambda$(\textbf{\emph{q}};\emph{v}) values at each phonon
vibration \emph{\textbf{v}} mode and at each phonon
\emph{\textbf{q}} momentum through Equs.~(\ref{equ4} and
\ref{equ5}), as compared in Fig. ~\ref{fig3}(a) and supplementary
Fig. S2~\cite{SM}. In a good agreement with previous experimental
observations\cite{be2009,be5}, the calculated ECFs at the 2D closed
circular \emph{k} momenta around the centered $\bar{\Gamma}$ are
anisotropic [\emph{i.e.}, ECF of No.1 mode in Fig.~\ref{fig3}(b)].
We have thus derived $\lambda^s_{e-ph}$ along the directions
rotating from the $\bar{\Gamma}$-$\bar{M}$ to
$\bar{\Gamma}$-$\bar{K}$ as shown in supplementary Fig.
S3~\cite{SM}, indicating the apparent anisotropy, again in nice
agreement with experimental findings [Fig.~\ref{fig3}(c)].
Strikingly, our calculations have evidenced that the DNSSs exhibit
the highly strong \emph{e-ph} coupling with the phonon modes, In
particular, at the phonon $\bar{K}$ momentum $\textbf{\emph{q}}$ =
(1/3,1/3,0)\cite{qk} (Fig.~\ref{fig3}(a)) their \emph{e-ph}
couplings are much stronger than those at all the other 34 phonon
\textbf{\emph{q}} momenta (corresponding to six
symmetry-inequivalent \textbf{\emph{q}} momenta, see supplementary
Fig. S3~\cite{SM}). In order to clearly visualize how the DNSSs
couple with these vibration modes, we have plot the mode-resolved
ECFs for nine selected vibration modes in the \emph{k} momenta of
the Be (0001) BZ at $\textbf{\emph{q}}$ = (1/3,1/3,0)\cite{qk},
including the aforementioned eight dominating No.1 $\sim$ No.6, No.
16 and No.29 modes [Fig.~\ref{fig3}(d to k)] and the lowest
contributed No.31 mode [Fig.~\ref{fig3}(l)]. For a sake of the
convenient comparison, the electronic Fermi surface is further plot
in Fig.~\ref{fig3}(b) with the aim of showing the exact \emph{k}
momenta where the DNSSs appear. On the one hand, a closed circle
surrounding the centered $\bar{\Gamma}$ point exactly corresponds to
the DNSSs, as marked by the SF-band1 in Fig.~\ref{fig1}(a) and, on
the other hand, three closed triangle-like electronic localized
states surrounding the centered $\bar{K}$ point are originated from
the topologically trivial surface states~\cite{be0} [SF-band2 and
SF-band3 in Fig.~\ref{fig1}(a)]. Remarkably, at each \emph{q}
momentum on the 2D circular \emph{k} momenta of the Fermi contour
where the topological protected DNSSs exactly appear in Fig.
~\ref{fig3}(c), we have also observed the highest bright closed
circular ECFs for all vibration modes, as shown for night selected
modes in Fig.~\ref{fig3}(e-m). However, at the triangle-like
\emph{k} momenta around $\bar{K}$ where the trivial surface
electronic states appear the brightness of the ECFs is much weak
than that of the centered circular ECFs around the $\bar{\Gamma}$.
It needs to be emphasized that the \emph{e-ph} couplings are
stronger for the eight surface localized phonon modes (No.1 $\sim$
No.6, No. 16 and No. 29 modes). Although the contributions of the
other modes to the \emph{e-ph} coupling are relatively not large,
the coupling between the DNSSs and these phonon modes are still
obvious. As evidenced in Fig.~\ref{fig3}(m) we have visualized the
ECFs of the No. 31 mode which exhibits the lowest contribution to
$\lambda_{e-ph}^s$, the coupled circular ECF can be clearly observed
as well. All these facts clearly evidence the robust strong
couplings of the DNSSs with the phonon vibration modes on the Be
(0001) surface, leading to its anomalously large surface \emph{e-ph}
coupling strength.

Furthermore, we have numerically calculated the weight of the
\emph{e-ph} coupling between the DNSSs and phonon modes to the total
$\lambda_{e-ph}^s$. Through Equ. (\ref{equ5}) at each phonon mode
and each phonon momentum, we have statistically derived ($i$) the
$\lambda(\emph{\textbf{q}};v)_{\textrm{DNSSs}}$ associated with the
electronic DNSSs by counting the decomposed ECFs over all electronic
$k$ momenta on the locations of the DNSSs, ($ii$) the
$\lambda(\emph{\textbf{q}};v)$ over all the electron \emph{k}
momenta in the surface BZ, and ($iii$) their ratio of
\begin{math} w =
\frac{\lambda(\emph{\textbf{q}};v)_{\textrm{DNSSs}}}{\lambda(\emph{\textbf{q}};v)}
\end{math}.
Among all the vibration \emph{v} modes and the phonon
\textbf{\emph{q}} momenta (see supplementary Table S1 and Table
S2~\cite{SM}), the lowest and highest ratios are $w$ = 77.6\% and
$w$ = 84.5\% originated from their \emph{e-ph} couplings with the
No.45 and No.3 phonon modes at $\textbf{\emph{q}}$ =
(1/3,1/3,0)~\cite{qk} in Fig.~\ref{fig3}(a), respectively. Of
course, the surface \emph{e-ph} coupling strength
$\lambda_{\textrm{DNSSs}}$ over all phonon momenta and all vibration
modes can be calculated by counting the ECFs at each $k$ momentum
where the DNSSs appear through supplementary Equs. (S1 and
S2)~\cite{SM}, consistently revealing that the DNSSs coupled
$\lambda_{\textrm{DNSSs}}$ is 80.2$\%$ of the total
$\lambda_{e-ph}^s$. As a result, these calculations confirm that the
anomalously large \emph{e-ph} coupling strength on the Be (0001)
surface stems from the DNSSs in coupling with phonon states.

Theoretically, it is still possible to predict the superconducting
$T_c$ of the Be (0001) surface with the theoretically derived
$\lambda_{e-ph}^s$ = 0.947 in combining with the ECF of
$\alpha^{2}$$F$($\omega$) in terms of the Dynes modified McMillan
formula \cite{TC1,TC2}. In a range of the effective screened Coulomb
repulsion constant $\mu$ = 0.10 $\sim$ 0.15, the superconducting
$T_c$ of the Be (0001) surface can be further estimated to be 15.3 K
$\sim$ 20.5 K. However, even down to 12 K the ARPES experiment
revealed no gap of the superconductivity~\cite{be4}. Yet, this
superconducting transition still needs to be confirmed, but it would
strictly depend on the high-quality samples of Be (0001) surface.

In summary, we have evidenced that the topologically protected
DNL-induced DNSSs lay a solid foundation for the novel enhancement
of the surface \emph{e-ph} coupling strength on the Be (0001)
surface. It would highlight a potential application correlated with
the \emph{e-ph} coupling interaction for various topological Dirac
nodal line semimetals.

\bigskip
\noindent {\bf Acknowledgments} The work was supported by the
National Science Fund for Distinguished Young Scholars (No.
51725103), by the National Natural Science Foundation of China
(Grant Nos. 51671193 and 51474202), and by the Science Challenging
Project No. TZ2016004. All calculations have been performed on the
high-performance computational cluster in the Shenyang National
University Science and Technology Park.

\noindent R. H. Li, and J. X. Li. contributed equally to this work.

\end{document}